\documentclass[a4paper]{jpconf}
\usepackage{graphicx}
\begin{document}
\title{Measurement of electrons from heavy-flavour decays in Pb-Pb collisions at  $\sqrt{s_{\mathit{\rm{NN}}}} = 2.76$ TeV with ALICE}

\author{Deepa Thomas for the ALICE collaboration}

\address{ERC-Research Group QGP-ALICE, Utrecht University, Princetonplein 5, 3584 CC Utrecht, The Netherlands}

\ead{deepa.thomas@cern.ch}

\begin{abstract}
The measurement of heavy-flavour (charm and beauty) production in ultra-relativistic heavy-ion collisions provides an important contribution to the study of the properties of the hot and dense medium created in such collisions. One approach to measure heavy-flavour production is via electrons from semi-leptonic decays of heavy-flavour hadrons. In this contribution we present the nuclear modification factor $(R_{{\rm{AA}}})$ and the azimuthal anisotropy $(v_{2})$ of heavy-flavour decay electrons in Pb-Pb collisions at $\sqrt{s_\mathit{{\rm{NN}}}}= 2.76$ TeV, which are sensitive to the interaction of c and b quarks with the medium. We also present, for the first time, the azimuthal angular correlations of heavy-flavour decay electrons and charged hadrons, $\Delta\phi(\mathit{HFE},h)$, in Pb-Pb collisions and the corresponding near-side yield ratio $(I_{\rm{AA}})$. 
\end{abstract}

\section{Introduction}
In collisions of lead ions at relativistic energies a high density, strongly interacting QCD medium is produced. Heavy-flavour particles produced in such collisions are a useful probe of the medium since they carry information about the full evolution of the system. One way to measure heavy-flavour production is through the measurement of electrons from heavy-flavour hadron decays $(\mathit{HFE})$. The observables discussed here reflect the interaction of heavy quarks with the medium. These are the nuclear modification factor $(R_{\rm{AA}})$, the elliptic flow of $(v_{2})$ of $\mathit{HFE}$ and the azimuthal angular correlations between $\mathit{HFE}$ and charged hadrons $(\Delta\phi(\mathit{HFE},h))$. $R_{\rm{AA}}$, which can be interpreted in terms of parton energy loss in the medium, is defined as the ratio of $p_{\rm{T}}$-differential yields measured in nucleus-nucleus collisions to the corresponding cross-section in pp collisions scaled by the nuclear overlap function $<T_{\rm{AA}}>$. 

Further insight into the transport properties of the medium can be obtained from the azimuthal anisotropy of particle production in non-central Pb-Pb collisions. The second harmonic coefficient of the Fourier expansion of particle azimuthal distributions, $v_{2}$, called elliptic flow, reflects the collective expansion of the medium at low $p_{\rm{T}}$ and the path-length dependence of parton energy loss. Further information on the parton fragmentation and energy loss can become accessible through azimuthal angular correlations $\Delta\phi(\mathit{HFE},h)$. 

\section{Dataset and data analysis}
Various detectors in the ALICE central barrel $(|\eta| < 0.9)$ are used for the $\mathit{HFE}$ measurements. The tracking system, which includes the Inner Tracking System (ITS) and the Time Projection Chamber (TPC) is used for the reconstruction of charged particle tracks and the measurement of their momenta. The particle identification system (PID) consists of a Time of Flight (TOF) detector, the TPC with which the specific ionization energy loss $(\rm{d}\mathit{E}/\rm{d}\mathit{x})$ of charged particles is measured and the Electromagnetic Calorimeter (EMCal), which registers electromagnetic showers in part of the central barrel acceptance ($|\eta| < 0.7$ and $80^{\circ} < \phi < 180^{\circ}$). At low $p_{\rm{T}}$  ($p_{\rm{T}}<6$ GeV$/c$ for the present analysis) electrons are identified with the TPC and TOF detectors, while with the EMCal the measurement can be extended to higher $p_{\rm{T}}$. In Pb-Pb collisions, the centrality was determined from the sum of the amplitudes of the signals in the VZERO detector and defined in terms of percentiles of the total hadronic Pb-Pb cross section. VZERO detector is also used to measure the reaction plane for the $v_{2}$ measurement. 

The results presented here are obtained from the Pb-Pb data at $\sqrt{s_{\mathit{\rm{NN}}}} = 2.76$ TeV from $2010$ and $2011$. The reference measurements used are from pp collisions at $7$ TeV. 

The general analysis strategy used for all measurements presented here involves the identification of electrons and the statistical removal of the background from sources other than heavy-flavour hadron decays $(\mathit{Non}$-$\mathit{HFE})$. This background is dominated by electrons from photon conversions in the detector material and Dalitz decays of light neutral mesons $($ mostly $\pi^{0}$ and $\eta)$. Two methods are available to remove the background, the invariant mass method and the cocktail method \cite{pp7HFE}. In the invariant mass method one makes use of the fact that electron-positron pairs from conversions and Dalitz decays have a small invariant mass and can thus be tagged. In the cocktail method, the $\mathit{Non}$-$\mathit{HFE}$ background is calculated with a  Monte-Carlo generator using the measured $\pi^{0}$ and $\eta$ $p_{\rm{T}}$-differential yields as input. 

\section{Results}
The nuclear modification factor is obtained from the $\mathit{HFE}$ yield measured in Pb-Pb collisions at $\sqrt{s_{\mathit{\rm{NN}}}} = 2.76$ TeV and using the corresponding cross-section from pp collisions at $\sqrt{s} = 7$ TeV to determine the reference for $p_{\rm{T}} < 8$ GeV$/c$ \cite{pp7HFE}. The latter is achieved by scaling the cross-section measured at $7$ TeV to $2.76$ TeV using a FONLL pQCD \cite{FONLL} based procedure \cite{scale}. For $p_{\rm{T}} > 8$ GeV$/c$ the result from the FONLL pQCD calculation at $\sqrt{s} =2.76$ TeV is used as reference. 

\begin{figure}[h]
\begin{minipage}[t]{0.5\linewidth}
\includegraphics[scale=0.35]{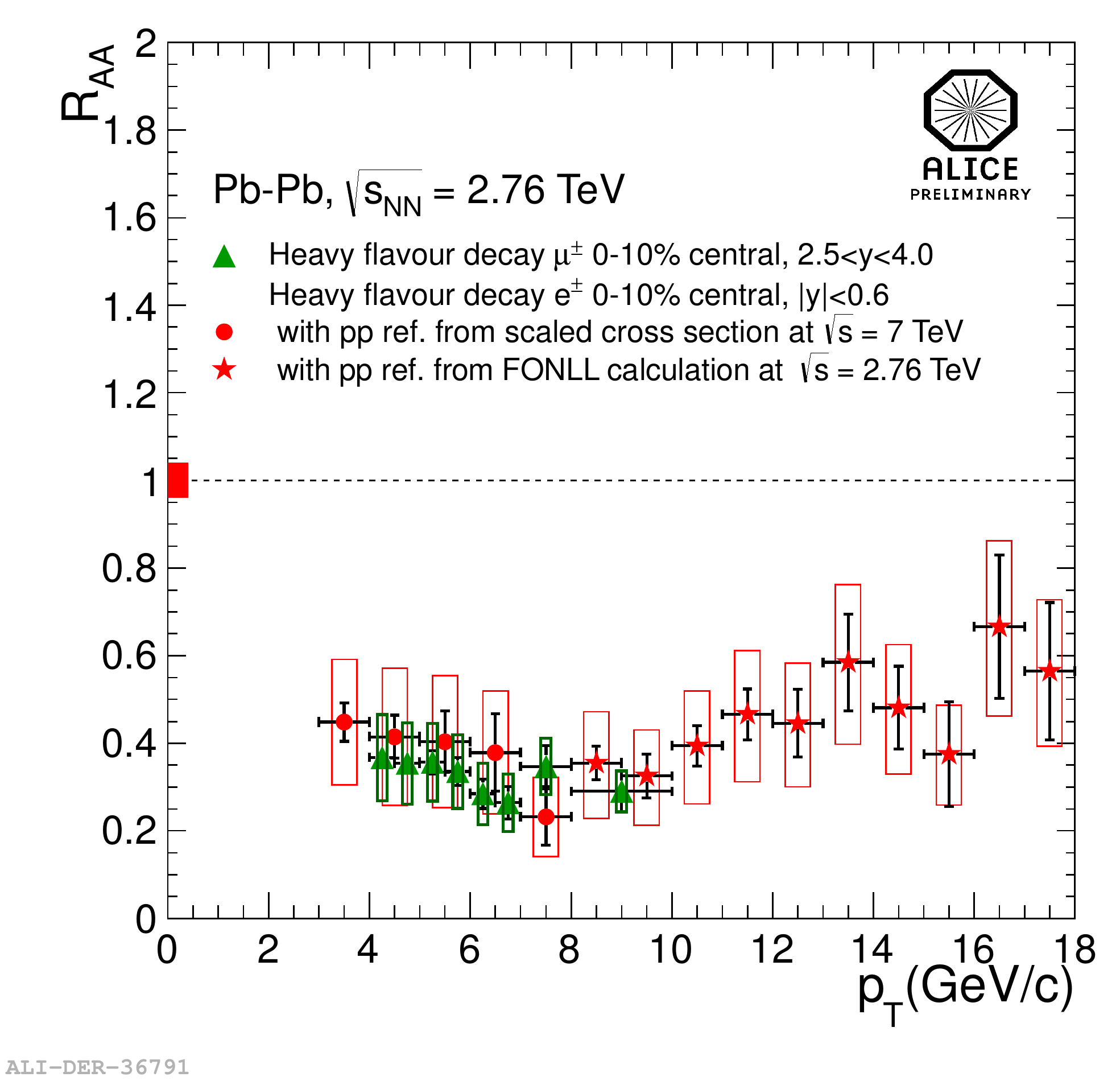}
\end{minipage}
\begin{minipage}[t]{0.5\linewidth}
\centering
\includegraphics[scale=0.35]{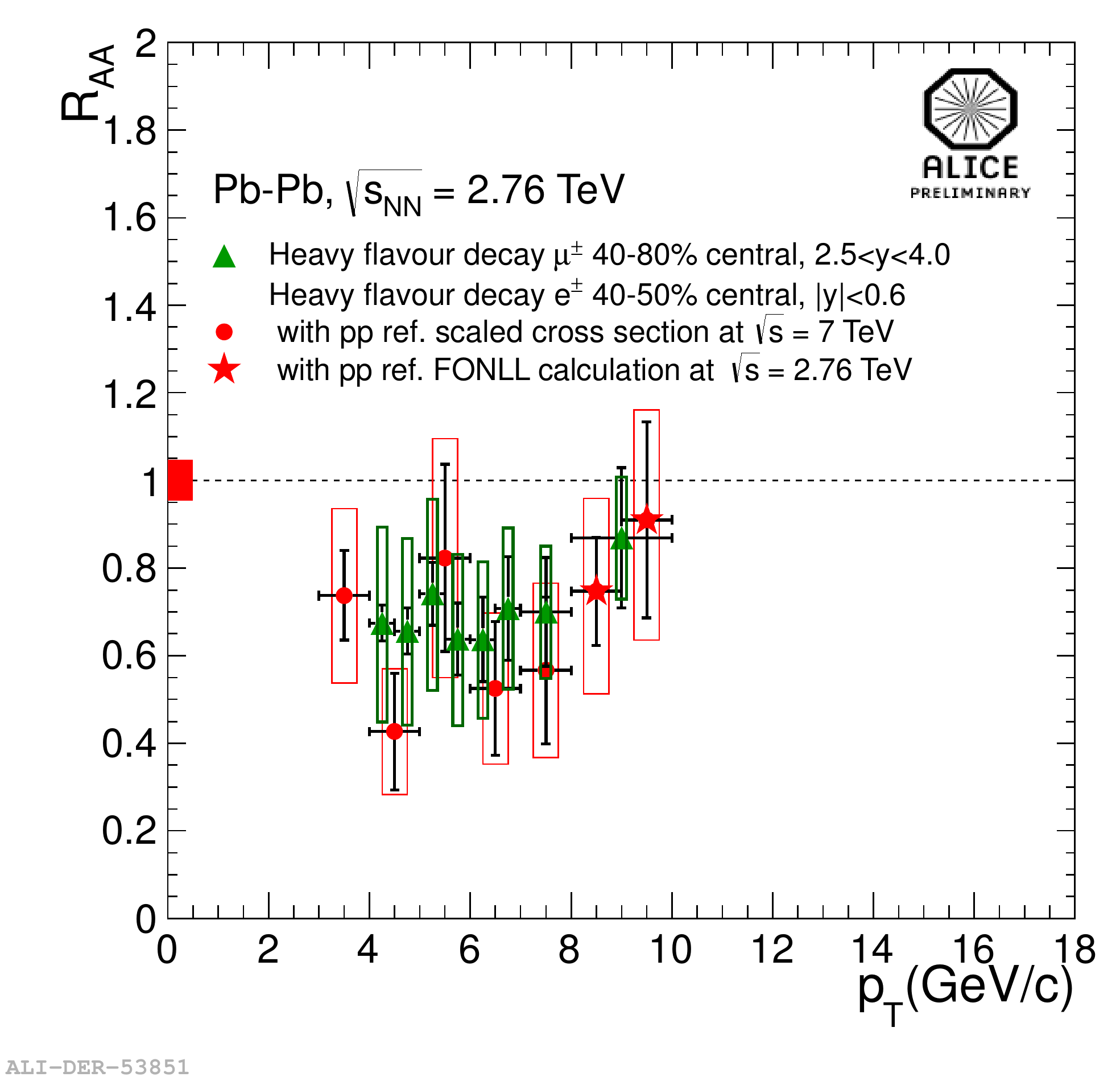}
\end{minipage}
\caption{\label{fig:RAA} $R_{\rm{AA}}$ of heavy-flavour decay electrons measured in the two centrality ranges $0-10\%$ (left) and $40-50\%$ (right) in Pb-Pb collisions at $\sqrt{s_{\mathit{\rm{NN}}}} = 2.76$ TeV, compared with the $R_{\rm{AA}}$ of heavy-flavour decay muons in the same centrality ranges.}
\end{figure}

The resulting $R_{\rm{AA}}$ of electrons from heavy-flavour hadron decays in central $(0-10\%)$ and semi-central $(40-50\%)$ Pb-Pb collisions is shown in Figure~\ref{fig:RAA}. A strong suppression of the HFE yield is observed in central collisions in the measured $p_{\rm{T}}$ range from $3$ to $18$ GeV$/c$. In semi-central collisions the suppression is less pronounced but still significant. The $\mathit{HFE}$ $R_{\rm{AA}}$ is consistent with that measured for heavy flavour decay muons at forward rapidity $(2.5<$y$<4)$ in both centralities.  

The elliptic flow of heavy-flavour decay electrons $(v_{2}^{\mathit{HFE}})$ is obtained using the $v_{2}$ measured for inclusive electrons $(v_{2}^{\mathit{Incle}})$ , the ratio $R_{\rm{NP}}$ of the $\mathit{HFE}$ and $\mathit{Non}$-$\mathit{HFE}$ yields $(N^{\mathit{HFE}}/N^{\mathit{Non}-\mathit{HFE}})$ and $v_{2}$ of $\mathit{Non}$-$\mathit{HFE}$ \cite{v2Phe}. 
\begin{equation}
v_{2}^{\mathit{HFE}} = \frac{(1+R_{\rm{NP}}) v_{2}^{\mathit{Incle}} - v_{2}^{\mathit{Non-HFE}}} {R_{\rm{NP}}} 
\label{Eq:flow}
\end{equation}

\begin{figure}[h]
\centering
\includegraphics[scale=0.65]{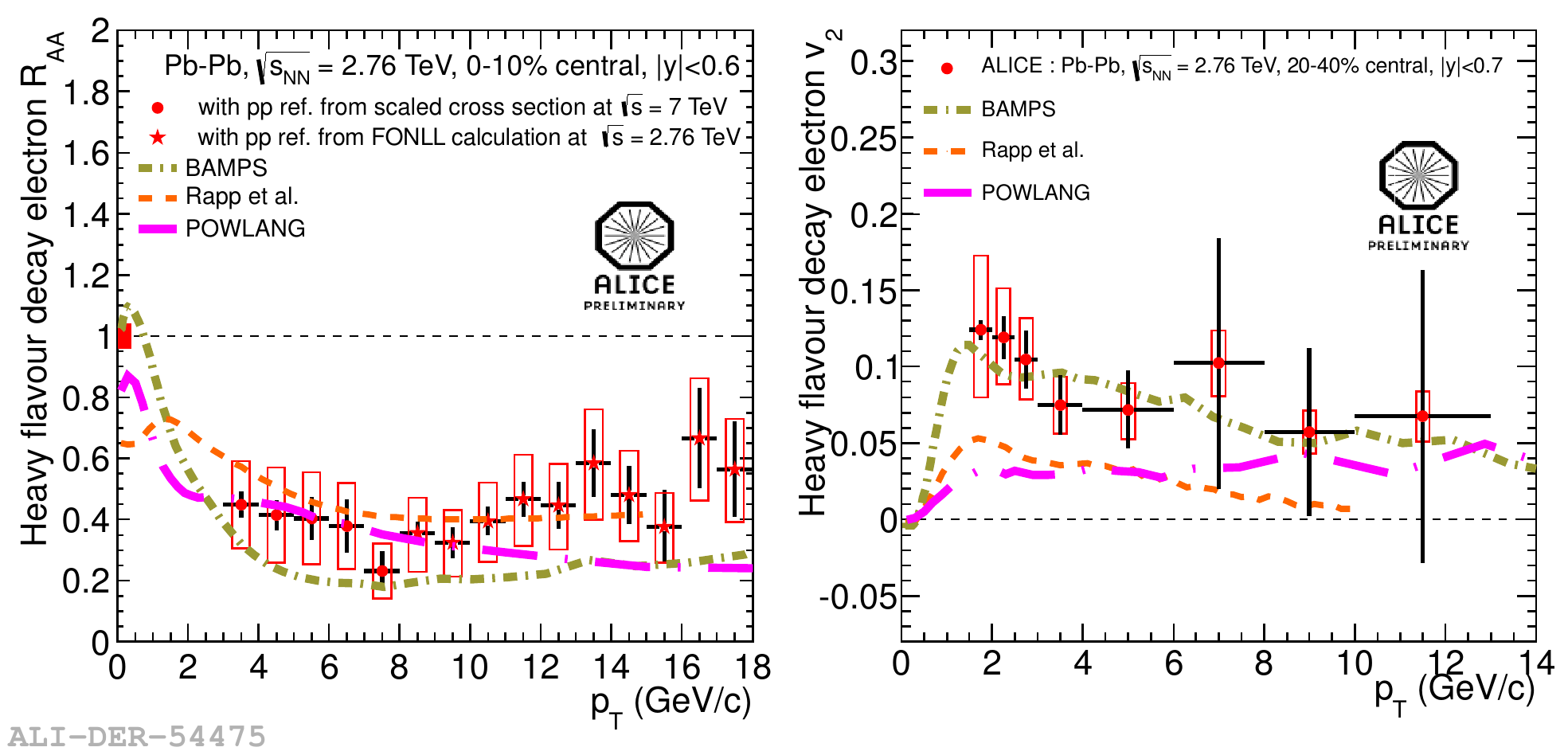}
\caption{\label{fig:Raav2Model} Comparison between measured $R_{\rm{AA}}$ and $v_{2}$ of heavy-flavour decay electrons and theoretical calculations \cite{bamps,rapp,powlang}.}
\end{figure}

The electron $v_{2}$ is obtained using the event plane method in which the particle azimuthal angle $(\phi)$ is measured relative to an estimator of the reaction plane determined from the azimuthal distribution of signals in the VZERO detector \cite{flowMet}. The background $v_{2}$ $(v_{2}^{\mathit{Non}-\mathit{HFE}})$ is calculated with a Monte-Carlo generator where the measured $v_{2}(p_{\rm{T}})$ of $\pi^{\pm}$, $\pi^{0}$ and $K^{\pm}$ are used as input, assuming $v_{2}^{\pi^{\pm}} = v_{2}^{\pi^{0}}$, $v_{2}^{\eta} = v_{2}^{K^{\pm}}$. The $v_{2}$ of direct photons is assumed to be zero. The $\mathit{HFE}$ $v_{2}$ is measured for $1.5 < p_{\rm{T}} < 13$ GeV$/c$, as shown in Figure~\ref{fig:Raav2Model}. A positive $v_{2}$ has been observed for $\mathit{HFE}$, which suggests the participation of heavy quarks in the collective flow at low $p_{\rm{T}}$. At high $p_{\rm{T}}$, it is more likely to be due to the path-length dependence of parton energy loss.

Various theoretical models predict $R_{\rm{AA}}$ and $v_{2}$ for $\mathit{HFE}$ as shown in Figure~\ref{fig:Raav2Model}. In BAMPS approach \cite{bamps} heavy quarks are transported through the  medium while undergoing collisional and radiative energy loss, Rapp et al \cite{rapp} include in-medium resonance scattering and coalescence of heavy quarks in the medium and in POWLANG \cite{powlang} heavy quarks are transported following a Langevin approach and considering collisional energy loss. BAMPS gives a good description of heavy-flavour decay electron $v_{2}$ but predicts a larger in-medium suppression than measured. The prediction from Rapp et al is consistent with the measured $R_{\rm{AA}}$, but it underestimates $v_{2}^{\mathit{HFE}}$. POWLANG describes the $\mathit{HFE}$ $R_{\rm{AA}}$ in the range $3 <p_{\rm{T}}^{e}<12$ GeV$/c$ but it underestimates $v_{2}^{\mathit{HFE}}$. The data provide a strong constraint on the theoretical description of heavy quark transport in the QGP.  

\begin{figure}[h]
\begin{minipage}[t]{0.6\linewidth}
\includegraphics[scale=0.53]{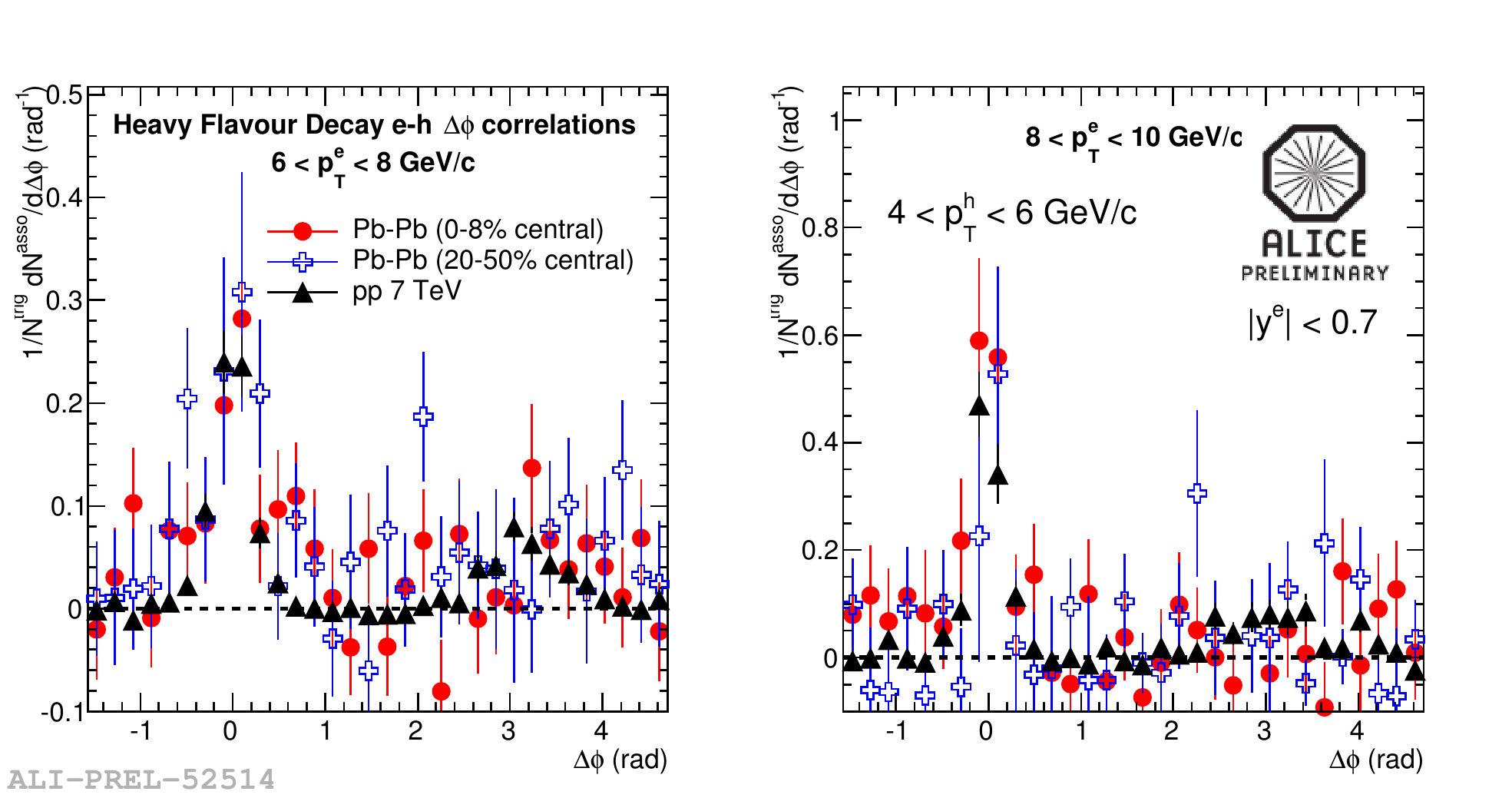}
\caption{\label{fig:deltaPhi} $\Delta\phi(\mathit{HFE},h)$ distribution for $4<p_{\rm{T}}^{h}<6$ GeV$/c$ in $0-8\%$ (red) and $20-50\%$ (blue) central Pb-Pb collisions at $\sqrt{s_{\mathit{\rm{NN}}}} = 2.76$ TeV compared to pp collisions at $\sqrt{s} = 7$ TeV (black).}
\end{minipage}\hspace{2pc}%
\begin{minipage}[t]{0.4\linewidth}
\centering
\includegraphics[scale=0.33]{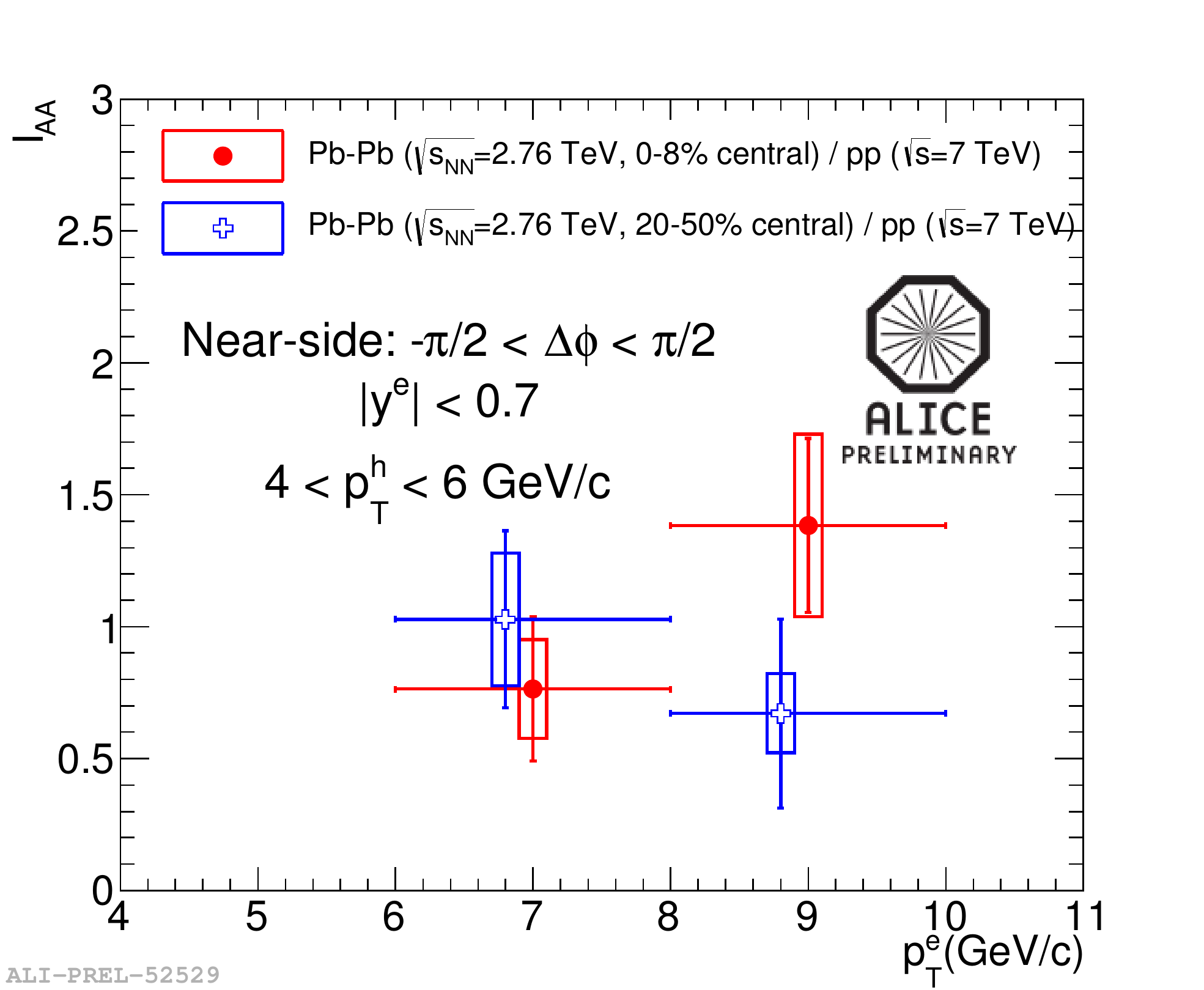}
\caption{\label{fig:IAA}$I_{\rm{AA}}$ for near side $(-\pi/2 < \Delta\phi < \pi/2)$ $\Delta\phi(\mathit{HFE},h)$ correlations for $4<p_{\rm{T}}^{h}<6$ GeV$/c$ in $0-8\%$ (red) and $20-50\%$ (blue) central Pb-Pb collisions at $\sqrt{s_{\mathit{\rm{NN}}}} = 2.76$ TeV.} 
\end{minipage}
\end{figure}

The $\Delta\phi(\mathit{HFE},h)$ correlation is measured for $4<p_{\rm{T}}^{h}<6$ GeV$/c$ in two trigger electron $p_{\rm{T}}$ intervals, $6<p_{\rm{T}}^{e}<8$ GeV$/c$ and $8<p_{\rm{T}}^{e}<10$ GeV$/c$ for two centralities, $0-10\%$ and $20-50\%$, in Pb-Pb collisions at $\sqrt{s_{\mathit{\rm{NN}}}} = 2.76$ TeV and in pp collisions at $7$ TeV. Since the vacuum fragmentation of heavy quarks is not expected to be different in pp collisions at $\sqrt{s}=7$ and $2.76$ TeV, the near-side yield in pp collisions at 7 TeV is used as reference. Figure~\ref{fig:deltaPhi} shows the $\Delta\phi(\mathit{HFE},h)$ in Pb-Pb and pp collisions after removing the pedestal defined between $1.5<\Delta\phi<2$ rad. To address a possible modification of the parton fragmentation function in the QCD medium, the near-side $(-\pi/2 < \Delta\phi < \pi/2)$ per-trigger yield is measured in Pb-Pb collisions and compared with the yield in pp collisions to obtain $I_{\rm{AA}}$, defined as ratio of the per-trigger yields in pp and Pb-Pb \cite{ALICEdihPap}. As depicted in Figure~\ref{fig:IAA}, the $I_{\rm{AA}}$ on the near-side is consistent with unity within the current uncertainties. 

\section{Conclusions}
With ALICE heavy-flavour production has been studied in Pb-Pb collisions via charm and beauty semi-leptonic decay channels to electrons. The nuclear modification factor, azimuthal anisotropy, and azimuthal angular correlations with hadrons have been measured for electrons from heavy-flavour hadron decays. $R_{\rm{AA}}$ in central Pb-Pb collisions demonstrates a significant energy loss of heavy quarks in the QCD medium. A non-zero $v_{2}^{\mathit{HFE}}$ has been measured for non-central Pb-Pb collisions suggesting participation of heavy quarks in the collective motion of the medium. First measurements of $\Delta\phi(\mathit{HFE},h)$ have been obtained for central and non-central Pb-Pb collisions. Higher statistics will help for more definite conclusions to be made in the future.

\end{document}